\documentclass[aps,prd,twocolumn,epsf,superscriptaddress]{revtex4}
\usepackage{graphicx}% Include figure files
\usepackage{amsmath,amssymb,latexsym}
\usepackage{bm}% bold math

\def\lbldef#1#2{\expandafter\gdef\csname #1\endcsname {#2}}

\def\href#1#2{#2}

\newcommand{\ber}{\begin{eqnarray}}
\newcommand{\eer}{\end{eqnarray}}

\newcommand{\beqar}{\begin{eqnarray}}

\newcommand{\eeqar}{\end{eqnarray}}
\newcommand{\dsl}
  {\kern.06em\hbox{\raise.15ex\hbox{$/$}\kern-.56em\hbox{$\partial$}}}

\newcommand{\eeqarr}{\end{eqnarray}}
\newcommand{\ZZ}{{\rm \kern 0.275em Z \kern -0.92em Z}\;}

\begin{document}

\preprint{
\hfil
\begin{minipage}[t]{3in}
\begin{flushright}
\vspace*{.4in}
NUB--xxx--Th--04\\
MIT--CTP--xxx\\
hep-ph/0410003
\end{flushright}
\end{minipage}
}

\title{Neutrinos as a Diagnostic of High Energy Astrophysical Processes}

\author{Luis A.~Anchordoqui}
\affiliation{Department of Physics,
Northeastern University, Boston, MA 02115
%\PRE{\vspace*{.1in}}
}

\author{Haim Goldberg}
\affiliation{Department of Physics,
Northeastern University, Boston, MA 02115
%\PRE{\vspace*{.1in}}
}

\author{Francis Halzen}
\affiliation{Department of Physics,
University of Wisconsin, Madison WI 53706
%\PRE{\vspace*{.1in}}
}

\author{Thomas J. Weiler}
\affiliation{Department of Physics and Astronomy,
Vanderbilt University, Nashville TN 37235
%\PRE{\vspace*{.1in}}
}
%\date{August 2004}
\begin{abstract}
A leading candidate for the extragalactic source of high energy
cosmic rays is the Fermi engine mechanism, in which protons
confined by magnetic fields are accelerated to very high energy
through repeated scattering by plasma shock fronts. In the process
of acceleration, collisions of trapped protons with the ambient
plasma produce pions which  decay to electromagnetic energy and
neutrinos. For optically thin sources, a strong connection between
the emerging cosmic rays and secondary neutrinos can be
established. In this context, we show the feasibility of using the
Glashow resonance as a discriminator between the $pp$ and
$p\gamma$ interactions in Fermi engines as sources of neutrinos.
In particular, we demonstrate how three years of observation at
the km$^3$ IceCube facility can serve as a filter for the
dominance of the $pp$ interaction at the source.

\end{abstract}

%\pacs{xxxx}

\maketitle

Neutrinos can serve as unique astronomical messengers. Except for
oscillations induced by transit in a vacuum higgs field, neutrinos
propagate without interactions between source and Earth, providing
powerful probes of high energy astrophysics~\cite{Gaisser:1994yf}.
The deployment of the km$^3$ IceCube facility at the South
Pole~\cite{Halzen:2003ve} will greatly increase the statistics
required for the realization of such a program.

The flavor composition of neutrinos originating at astrophysical
sources can also serve as a probe of new physics in the
electroweak sector~\cite{Beacom:2002vi}. Specifically,
IceCube  will have the capability to clearly
identify neutrino species~\cite{Beacom:2003nh}, and consequently will
be able to measure deviations of flavor composition from standard
expectations. In addition, in resonant scattering $\overline \nu_e
e^- \to W^- \to {\rm anything}$~\cite{Glashow:W}, the detector can
simultaneously discriminate between $\nu_e $ and $\overline \nu_e.
$ In this Letter, we show that the signal for
$\overline \nu_e$ at the Glashow resonance, when normalized to the
total $\nu + \overline \nu$ flux, can be used to differentiate
between  the two primary candidates ($p\gamma$ and $pp$ collisions)
for neutrino-producing interactions in optically thin sources of cosmic rays.

It is helpful to envision the cosmic ray engines as machines where
protons are accelerated and (possibly) permanently confined by
the magnetic fields of the acceleration region. The production of
neutrons and charged pions and subsequent decay produces both neutrinos and cosmic
rays: the former via $\pi^+ \rightarrow e^+ \nu_e \nu_\mu \overline
\nu_\mu$ (and the conjugate process -- more on this below), the latter via
neutron diffusion from the region of the confined protons. If the
neutrino-emitting source  also produces high energy cosmic rays,
then  pion production must be the principal agent for the high
energy cutoff on the proton spectrum~\cite{Biermann:1987ep,note1}.
Conversely, since the protons must undergo sufficient
acceleration, inelastic pion production needs to be small below
the cutoff energy; consequently,  the plasma must be optically
thin. Since the interaction time for protons is greatly increased
over that of neutrons because of magnetic confinement, the
neutrons escape before interacting, and on decay give rise to the
observed cosmic ray flux. A desirable property of this low-damping
scenario is that a single source will produce cosmic rays with a smooth
spectrum across a wide range of energy.

For optically thin sources, the neutrino power density scales
linearly with the cosmic ray power density $\dot \epsilon_{_{\rm
CR}}$~\cite{Waxman:1998yy}. The actual value of the neutrino flux
depends on what fraction of the proton energy is converted to
pions (which then decay to neutrinos and photons). To quantify
this, here we define $\epsilon_\pi$ as the ratio of pion energy to
the emerging nucleon energy at the source. Since there is about
one cosmic ray (neutron) produced per proton collision, a
significant consequence of the neutron leakage model is that
$\epsilon_\pi$ is expected to be approximately equal to the
fraction of collision energy carried off by pions in a single
collision. Following Waxman and Bahcall~\cite{Waxman:1998yy}, the
expected total (all species) neutrino flux is
\begin{equation}
E_\nu^2 \,\Phi_{\nu + \overline \nu}^{\rm total} \approx 2 \times
10^{-8} \,\, \epsilon_\pi\, \xi_z\, \, {\rm GeV}\,\, {\rm
cm}^{-2}\,\, {\rm s}^{-1} \,\, {\rm sr}^{-1}\,\, ,
\label{wb}
\end{equation}
where for no source evolution $\xi_z \approx 0.6,$ whereas for an
evolution $\propto (1+z)^3$ as seen in the star-formation rate,
$\xi_z \approx 3$~\cite{ref}.

Depending on the relative ambient gas and photon densities, pion
production proceeds either through inelastic $pp$
scattering~\cite{Anchordoqui:2004eu}, or photopion production
predominantly  through the resonant  process $p\gamma \rightarrow
\Delta^+\rightarrow n\pi^+\,{\rm
or}\,p\pi^0$~\cite{Waxman:1998yy}. For the first of these, pion
production is well-characterized by a central rapidity (Feynman)
plateau, yielding $\epsilon_\pi\approx
0.6$~\cite{Alvarez-Muniz:2002ne}. For resonant
photoproduction, the inelasticity is kinematically determined by
requiring equal boosts for the decay products of the
$\Delta^+$~\cite{Stecker:1968uc}, giving $\epsilon_\pi\approx
0.25$~\cite{note2}.

The first production mechanism (described above) involves the
dominance of inelastic $pp$ collisions in generating charged
pions. The nearly isotopically neutral mix of pions will create on decay
a neutrino population in the ratio
$N_{\nu_\mu}=N_{\overline\nu_\mu} =
2N_{\nu_e}=2N_{\overline\nu_e}.$ In propagation to Earth, a
distance longer than all oscillation lengths, flavor-changing
amplitudes are replaced by probabilities~\cite{Eidelman:2004wy},
resulting in equal fluxes for all six neutrino species, so that
$N^{\rm Earth}_{\overline\nu_e}=\frac{1}{6}\ N_{\nu + \overline
\nu}^{\rm total}.$ Here, we have incorporated maximal $\nu_{\mu}
\leftrightarrow \nu_{\tau}$ mixing and  the known smallness of
$|\langle \nu_e|\nu_3\rangle|^2$,  where $\nu_3 \simeq
(\nu_\mu+\nu_\tau)/\sqrt{2}$ is the third neutrino eigenstate.

The second mechanism proceeds via photoproduction of pions by
trapped protons on the thermal photon background, leaving the
isotopically asymmetric process $p\gamma\rightarrow
\Delta^+\rightarrow \pi^+ n$ as the dominant source of neutrinos.
At production, $N_{\nu_\mu}=N_{\overline\nu_\mu} = N_{\nu_e} \gg
N_{\overline\nu_e}.$ After oscillation, the $\overline\nu_e$ flux
at Earth is
\begin{eqnarray}
N^{\rm Earth}_{\overline\nu_e} &=& N_{\overline\nu_\mu} \,
P(\overline\nu_\mu\rightarrow\overline\nu_e) \nonumber \\
&=& \frac{1}{3}\ \sin^2{\theta_\odot}\,\, \cos^2{\theta_\odot}\,\, N_{\nu
+ \overline \nu}^{\rm total}\ \ .
\end{eqnarray}
Using the most recent SNO value of the solar mixing angle
$\theta_\odot \simeq 32.5^\circ$~\cite{Ahmed:2003kj}, we
obtain $N^{\rm Earth}_{\overline\nu_e}=\frac{1}{15}\ N_{\nu +
\overline \nu}^{\rm total}.$ Because of averaging over various
thermal environments, this relation  holds across the neutrino
spectrum~\cite{Waxman:1998yy}.

The preceding discussion indicates the feasibility of
differentiating between the two processes as follows: {\em (i)}
obtain a normalization for the total flux from data outside the
Glashow resonance region; {\em (ii)} measure the event rate in the
resonance region, and compare with the ratios given above. In what
follows, we will assess the potential of IceCube to obtain a
statistically significant signal-to-background ratio.

The cross section for Glashow resonant scattering reads
\begin{equation}
\sigma = \frac{\pi}{2 m_e \, E_\nu}\, |\mathfrak{M}|^2\,
\delta(2 m_e E_{\nu} - m_W^2) \,\,,
\end{equation}
where $|\mathfrak{M}|^2 = g^2 m^2_W/2,$ $m_e$ and $m_W$ are the
electron and $W$ masses, and $g^2 = 4 \pi \alpha (m_W)/\sin^2
\theta_{\rm W} (m_W) \simeq 0.43.$ Now, for an incoming flux of
antineutrinos $J_{\overline \nu_e}(E_\nu)$ the event rate
(assuming 100\% detection efficiency) reads
\begin{eqnarray}
\frac{d{\cal N}}{dt} & = & N_{\rm eff} \,\,\Delta \Omega \,\,
\int dE_\nu \,J_{\overline \nu_e}(E_\nu) \,\, \sigma(E_\nu) \nonumber\\
         & = & N_{\rm eff} \,\, \Delta\Omega \,\,
 \frac{\pi g^2}{4 m_e} \,\, J_{\overline \nu_e} \left(\frac{m_W^2}{2 m_e}\right)
\,\,,
\end{eqnarray}
where $N_{\rm eff}\simeq 6\times 10^{38}$ is the number of target
electrons in the effective volume $V \sim 2$~km$^3$ of the IceCube
experiment, and $\Delta\Omega \simeq 2\pi$ is the solid angle
aperture~\cite{note3}. If neutrinos are produced via $pp$
inelastic collisions, according to the previous discussion
(with $\epsilon_\pi=0.6$ and $\xi_z=3$), the antineutrino flux on
Earth is $E_\nu^2\ J_{\overline \nu_e}(E_\nu)=6\times 10^{-9}$~GeV
cm$^{-2}$ s$^{-1}$ sr$^{-1}$, so that the event rate is $d{\cal
N}/dt =4.6$ yr$^{-1}.$ For antineutrinos produced in photopion
interactions ($\epsilon_\pi=0.25$), the expected rate is a factor of
6 smaller, $d{\cal N}/dt =0.8$~yr$^{-1}.$

Can the expected resonant signal be differentiated from the continuum
background? In order to estimate  the background, we note that
all but the muon
decays of the $W$ devolve into an electromagnetic shower in the
detector. For the hadronic decays (which constitute 70\% of the
total), the interaction mean free path of charged pions in ice is orders of
magnitude smaller than the pion decay length at TeV energies,
allowing a progressive channeling of  nearly all the energy into
electromagnetic modes through $\pi^0$ decay. For the $e^-\overline
\nu_e$ and $\tau^-\overline \nu_\tau$ decay modes, there will be a
single electromagnetic burst triggered by the electron interaction
or by the $\tau$ decay, each however manifesting only half the
energy. As a conservative procedure, we will  use in our signal
estimate only the hadronic events with full energy in the resonant
band, giving  (in the case of $pp$ interactions) a signal $d{\cal
N}/dt =3.2$ yr$^{-1}.$ The background will then constitute the
non-resonant scattering of $\nu_e$ and $\overline \nu_e$ within
the resonant band. This is calculated as an integral of the flux
times the charged current cross section~\cite{Gandhi:1998ri} over
the resonant acceptance bin $(10^{6.7}~{\rm GeV},\, 10^{6.9}~{\rm
GeV})$ of the AMANDA detector~\cite{Ackermann:2004zw}. The process
being off-resonance, the Earth is transparent and so we take
$\Delta\Omega =4\pi$. The number of
target nucleons is twice the number of electrons. This procedure
yields a background of 0.6 events yr$^{-1}$. Thus, within 3 years
of data accumulation (1/5 of the total lifetime of the experiment),
we expect about 2 background events, with a
signal of 9.6 events, well above the 99\%CL for discovery
(corresponding to 6.69 events~\cite{Feldman:1997qc}).

Electron antineutrinos can also be produced through neutron
$\beta$-decay. This contribution turns out to be negligible. To
obtain an estimate, we sum over the neutron-emitting sources out
to the edge of the universe at a distance
$1/H_0$~\cite{Anchordoqui:2003vc}
\begin{equation}
\Phi_{\overline \nu_e} = \frac{m_n}{8\,\pi \,\epsilon_0\, H_0}
\int_{\frac{m_n E_{\overline \nu}}{2 \epsilon_0}}^{E_n^{\rm max}}
\frac{dE_n}{E_n} \,\, {\cal F}_n (E_n) \,\, ,
\end{equation}
where  ${\cal F}_n (E_n) $ is the neutron volume emissivity and
$m_n$ the neutron mass. Here, we have assumed that the neutrino is
produced monoenergetically in the neutron rest frame, i.e.,
$\epsilon_0 \sim \delta m_N (1 - m_e^2/\delta^2 m_N)/2 \sim
0.55$~MeV, where $\delta m_N \simeq 1.30$~MeV is the neutron
proton mass difference. An upper limit can be placed on ${\cal
F}_n$ by assuming an extreme case in which all the cosmic rays
escaping the source are neutrons, i.e.,
$\dot \epsilon_{_{\rm CR}} = \int dE_n\,\, E_n\,\, {\cal F}_n(E_n).$
With the production rate of ultrahigh energy protons
$\dot \epsilon_{_{\rm CR}}^{[10^{10}, 10^{12}]} \sim 5 \times
10^{44}$~erg Mpc$^{-3}$ yr$^{-1}$~\cite{Waxman:1995dg}, and an assumed
injection spectrum ${\cal F}_n\propto E_n^{-2},$ we obtain
\begin{equation}
E_{\nu}^2\ \Phi_{\overline \nu_e}\approx 3\times 10^{-11} \,\,
{\rm GeV} {\rm cm}^{-2} \, {\rm s}^{-1} \,
{\rm sr}^{-1}\,\,,
\end{equation}
about 3 orders of magnitude smaller than the charge pion
contribution with $\epsilon_\pi=0.6.$ (Note that oscillations will
reduce the  $\overline \nu_e$ flux on Earth  by 40\% from this
value.)

Finally, we note that in the case that pions are produced via
$p\gamma\rightarrow \Delta^+ \rightarrow  N \pi$, the event rate for the
Glashow resonance cannot
be separated from background. Thus, the presence or absence of a
signal at the Glashow resonance can be used as a filter for
$pp$-dominance at the source.

In summary, we have shown how several years of data collection at
IceCube can potentially isolate the dominance of inelastic $pp$
interactions as neutrino progenitors in cosmic ray sources. The
analysis depends on the  sources being optically thin, which can
be ascertained if the observed diffuse neutrino flux is in
agreement with Eq.~(\ref{wb}). Then a signal at the Glashow resonance will
provide a strong marker for $pp$-dominance.
 \\

\underline {Note added:} 
The previous version of this paper contained a brief discussion of how
the flux of neutrinos from optically thin
sources can be used as a marker of the cosmic ray
Galactic/extra-galactic transition. This topic has received
detailed  discussion in a separate paper~\cite{Ahlers:2005sn}.\\

This work has been partially supported by US NSF (Grant Nos. OPP-0236449,
PHY-0140407, PHY-0244507), US DoE (Grant Nos. DE-FG02-95ER40896, DE-FG05-85ER40226),
NASA-ATP 02-000-0151, and the Wisconsin Alumni Research Foundation.\\


\begin{thebibliography}{99}
\bibitem{note1} For
proton energies $<10^{11}$ GeV, the Hillas criterion is easily
satisfied.
\bibitem{ref}The assumption of optically
thin sources also guarantees that the muon decays well before it
interacts. Note that the condition on neutron leakage requires
that the neutron interaction length is larger than the
accelerating region. Since the muon interaction length (mostly
from pair production and bremsstrahlung)is in turn larger than
that of the neutron,, the muon will decay without losing energy,
either inside or outside the source, giving rise to secondary
neutrinos, each with 1/4 of the pion energy.
\bibitem{note3} Note that, off-resonance, the IceCube detector will have nearly full solid
angle acceptance at PeV energies.
\end{thebibliography}
\end{document}